\documentclass[prl,twocolumn,showpacs,superscriptaddress,floatfix,lengthcheck]{revtex4}
\usepackage{graphicx}
\usepackage{dcolumn}
\usepackage{bm}
\usepackage[usenames]{color}

\begin{document}
\title{Proposal for manipulating and detecting spin and orbital states of trapped electrons on helium using cavity quantum electrodynamics}
    \author{D. I. Schuster}
    \affiliation{Department of Applied Physics and Physics, Yale University}
    \author{A. Fragner}
    \affiliation{Department of Applied Physics and Physics, Yale University}
    \author{M. I. Dykman}
    \affiliation{Department of Physics and Astronomy, Michigan State University}
    \author{S. A. Lyon}
    \affiliation{Department of Electrical Engineering, Princeton University}
    \author{R. J. Schoelkopf}
    \affiliation{Department of Applied Physics and Physics, Yale University}
\date{\today}

\begin{abstract}
We propose a hybrid architecture in which an on-chip high finesse superconducting cavity is coupled to the lateral motion and spin state of a single electron trapped on the surface of superfluid helium.  We estimate the motional coherence times to exceed $15\, \rm{\mu s}$,  while energy will be coherently exchanged with the cavity photons in less than $10$ ns for charge states and faster than $ 1\mu$s for spin states, making the system attractive for quantum information processing and strong coupling cavity quantum electrodynamics experiments. The cavity is used for non-destructive readout and as a quantum bus mediating interactions between distant electrons or an electron and a superconducting qubit. 
\end{abstract}

\maketitle

The field of experimental quantum information processing has made significant progress in recent years. Many different physical implementations are being actively explored, including trapped ions~\cite{leibfried_experimental_2003,haffner_scalable_2005}, semiconductor quantum dots~\cite{petta_coherent_2005,koppens_driven_2006}, and superconducting qubits~\cite{steffen_measurement_2006,dicarlo_demonstration_2009}.  In particular, the strong coupling to microwave photons possible in circuit quantum electrodynamics (QED) architectures~\cite{wallraff_strong_2004} has sparked interest in creating hybrid quantum systems capable of combining the advantages of different qubit implementations. In these proposals, a superconducting transmission line cavity acts as an interface between superconducting circuits and microscopic quantum systems, such as polar molecules~\cite{andre_coherent_2006,rabl_hybrid_2006}, electron spins~\cite{imamoglu_cavity_2009,wesenberg_quantum_2009} or ultra-cold atoms~\cite{verdu_strong_2009}, typically with smaller couplings but much better coherence than superconducting qubits.   Single electrons trapped above the surface of superfluid helium~\cite{Andrei1997} might play a unique role as they can independently form a strongly coupled cavity QED system or act in concert with superconducting qubits.

Interest in electrons on helium is motivated in part by their exceptional properties, including the highest measured electron mobility~\cite{shirahama_surface_1995} and long predicted spin coherence times~\cite{lyon_spin-based_2006}.  For these reasons the system was used in one of the first quantum information processing proposals~\cite{platzman_quantum_1999}.  The initial proposal focused on the motional states of a single trapped electron \textit{normal} to the helium surface~\cite{dykman_qubits_2003}, which promise long coherence times but have transition frequencies in the inconvenient range of $100$~GHz.  Further, the electrons were to be detected destructively.  More recently it has been proposed to use electron spins~\cite{lyon_spin-based_2006}, and the possibility of moving electrons at MHz rates was demonstrated~\cite{sabouret_signal_2008}, but it was not clear how to best read-out or couple such spin states.  

Here, we address these challenges using the circuit QED architecture~\cite{wallraff_strong_2004}, and show that both the electron's motion and spin can be used to reach the strong coupling limit of cavity QED, where the coupling between the electron and cavity is larger than their decoherence rates, allowing for a wide variety of quantum optics and quantum information experiments.  The quantized in-plane motion, \textit{parallel to the helium surface}, can be engineered to have transition frequencies of a few GHz and is readily coupled to an on-chip cavity for non-destructive readout analogous to that used for superconducting qubits~\cite{wallraff_strong_2004} or electron cyclotron motion in g-2 experiments~\cite{Peil1999}.  The cavity also mediates interactions between individually trapped electrons allowing for multi-qubit gates similar to those demonstrated in superconducting systems~\cite{dicarlo_demonstration_2009}.  In addition, the spin-photon coupling would be significantly enhanced by a controllable spin-orbit coupling.

The trapped electrons can be considered as quantum dots on helium operating in the single electron regime. These dots would be sufficiently small (sub-micron) that the lateral spatial confinement and potential depth will determine the orbital properties.    A Jaynes-Cummings coupling between in-plane states and out-of-plane states in such dots was proposed recently~\cite{zhang_jaynes-cummings_2009}. The feasibility of creating such nano-scale traps is buoyed by a recent experiment which has detected single electron tunneling events~\cite{papageorgiou_counting_2005}.      However, so far there have been no observations of either intradot quantization or spin resonance on helium.

It is instructive to compare electrons on helium with semiconductor quantum dots.  In most traditional two dimensional electron gases (2DEG's) such as in GaAs, the electrons form a degenerate gas with small effective masses, renormalized g-factors, and strong interactions with the lattice.  In particular the strong piezoelectric coupling leads to short coherence times for the motional states ($\sim100$~ps)~\cite{fujisawa_spontaneous_1998}.  For this reason spin is typically used~\cite{petta_coherent_2005,koppens_driven_2006}, but its coherence time can be strongly affected by nuclear spins~\cite{johnson_triplet-singlet_2005}.  In contrast, electrons on helium form a 2DEG at the interface between vacuum and superfluid, retaining their bare mass and g-factor.  With the techniques described here, single electron quantum dots on helium promise some advantages over traditional semiconducting dots.   We predict the decay of the orbital states to be $10^6$ times slower than in GaAs.  Further, superfluid $^4$He has no nuclear spins ($10^{-6}$ natural abundance of $^3$He), leading to long predicted spin coherence times\cite{lyon_spin-based_2006}, which are primarily limited by current noise in the trap leads.   Perhaps most importantly, electrons on helium is a fascinating system where coherent single particle motion has not been accessible until now.

An electron near the surface of liquid helium experiences a potential due to the induced image charge of the form $V = -\Lambda /z$, with $\Lambda=e^2(\epsilon-1)/4(\epsilon+1)$ and $\epsilon \approx 1.057$.  Together with the 1 eV barrier for penetration into the liquid, the image potential results in a hydrogen-like spectrum $E_{\rm{n}}=-R/n^2$ of motion normal to the surface, with effective Rydberg energy $R \sim 8$ K and Bohr radius $8$ nm~\cite{Andrei1997}.  At the working temperature of $50$ mK the electron will be frozen into the ground out-of-plane state, and the helium will be a superfluid with negligible vapor pressure.

\begin{figure}
\centering
\includegraphics{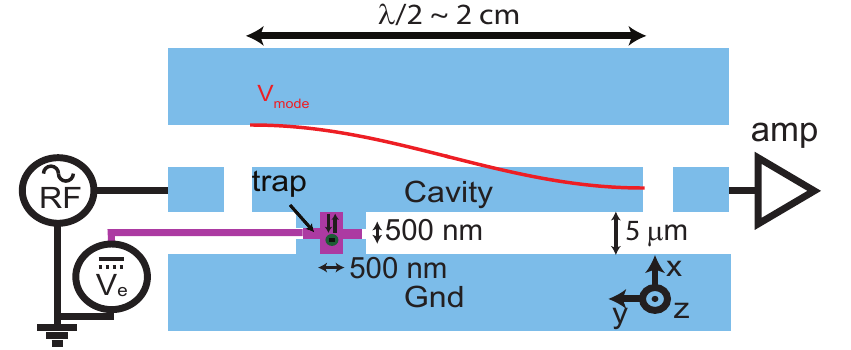}
\caption[Electron trap embedded in CPW]{(color) Top view of electrostatic electron trap.  The ground plane and cavity center pin are shown in blue, while the trap electrode is magenta.  The configuration of center pin and ground plane provide two-dimensional confinement.  A DC voltage, $V_{\rm{e}}$ is provided via a wire insulated from the resonator.   Manipulation and readout is performed via an RF voltage applied to the input port of the resonator with the modified signal measured by a cryogenic amplifier at the output port.}\label{fig:TrapTopView}
\end{figure}

\begin{figure}
\centering
\includegraphics{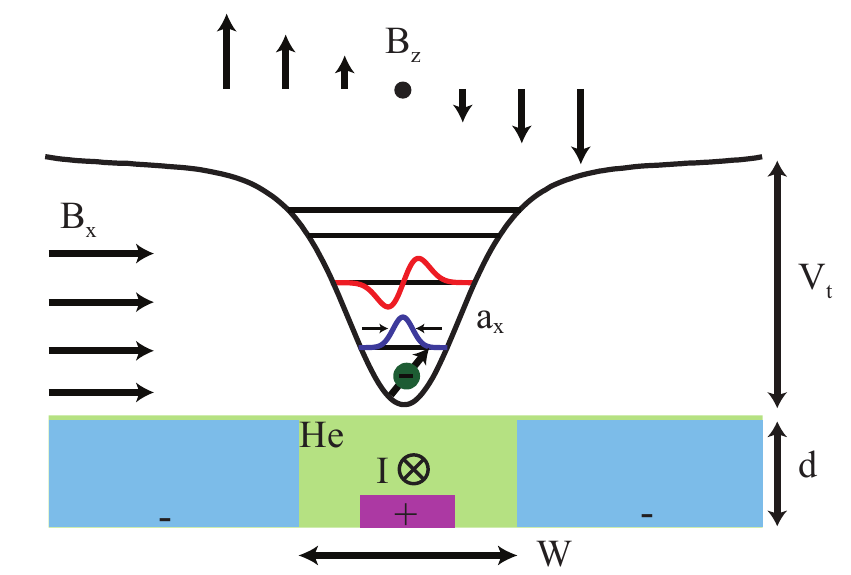}
\caption[Energy levels of trapped electron]{(color) Side view of trap electrodes with energy levels and wavefunctions of electron motional state.  The electron is confined to the surface of the helium film of thickness, $d$.  The trap electrode (magenta) is biased positive relative to the ground and center pin (blue) of the CPW to laterally confine the electron.  These electrodes form a confining potential which is harmonic to first order, but which flattens over the outer electrodes, giving it a small softening anharmonicity.  A sample potential and nearly-harmonic wavefunctions are shown.  The spatial extent of the electron zero-point motion $a_x$ is small compared to the characteristic size of the trap $w$.  To define a spin quantization axis a magnetic field in the x-direction is applied.  To couple the motional and spin degrees of freedom current is sent through the center electrode creating a z-field gradient within the trap.}\label{fig:MotionalLevels}
\end{figure}

With the vertical motion eliminated, the electron's lateral motion within an electrostatic trap could be coupled to the electric field of a superconducting transmission line cavity.  As shown in Fig.~\ref{fig:MotionalLevels}, the cavity center-pin and ground plane form a split-guard ring around a positively biased trap electrode.  We approximate the trapping potential in each of the lateral dimensions as being nearly parabolic, with level spacing $\approx \hbar \omega_{x,y}$.  We assume a single electron in a high-aspect ratio trap so that the $x$ and $y$ motional frequencies are distinct, with $\omega_x<\omega_y$. The Hamiltonian of the electron near the potential minimum can be approximated as
\begin{equation}
H_e=\frac{\hat{p_x}^2}{2 m_e} + \frac{1}{2} m_e \omega_x^2 \hat{x}^2 + \hbar \alpha \frac{\hat{x}^4}{3 a_x^4}
\end{equation}
Here, $a_x=(\hbar/ m_e \omega_{x})^{1/2}$ is the standard deviation of the motional ground state wavefunction and $\alpha$ is the anharmonicity. Because the trap is small and the potential must flatten at the outer electrodes, $\alpha< 0$. The $n$ to $n+1$ transition frequency is $\omega_{x,n}\approx \omega_{x,0}+(n+1) \alpha$.  The electron motion can be treated as a qubit when $|\alpha|$ is larger than the decoherence rates.  The scaling of the system parameters with geometry (see Fig.~\ref{fig:MotionalLevels}) can be estimated analytically by approximating the trap potential as $V_t \cos (2\pi x /W)$.  In this case $\omega_x = 2\pi (e V_t / m_e W^2)^{1/2}$, $\alpha = (2\pi/W)^2 \hbar/8m_e $, and $V_t \approx V_e e^{-2\pi d / W}$.  Therefore one can tune the motional frequency by adjusting the bias voltage, determine the anharmonicity by the trap size (confinement effects), and trade-off sensitivity in bias voltage for sensitivity to trap height (generally $d \sim W$ so as to avoid exponential sensitivity to film
thickness). 
   
The microwave environment and the trapping potential are simulated using Sonnet\textregistered and  Maxwell\textregistered, respectively and then Schr\"odinger's equation is numerically solved to find the resulting wavefunctions for the geometry shown in Figs.~\ref{fig:TrapTopView} \& \ref{fig:MotionalLevels}.  Using physically reasonable trapping parameters: helium depth $d=500 \, \rm{nm}$, trap size $W=500\, \rm{nm}$, trapping voltage $V_{\rm{e}}=10\, \rm{mV}$ results in a trap depth $e V_{\rm{t}}/h\approx20\, \rm{GHz} \gg k_{\rm{B}} T$ deep enough to prevent thermal escape, and a transition frequency $\omega_{x}/2\pi \approx 5\, \rm{GHz}$ convenient to microwave electronics.  The cavity can be represented by the Hamiltonian $H_r = \hbar \omega_r (a^{\dag} a+1/2)$, with $\omega_r/2\pi \approx 5\, \rm{GHz}$ close to the desired motional frequency.  The electron's motion within the trap is affected by and induces an electric field in the microwave cavity.  If the level spacing $\hbar \omega_x$ is in resonance with the energy of a cavity photon $\hbar \omega_r$, the two systems can exchange energy at the vacuum Rabi frequency, $2 g = \sqrt{2}e a_x E_0/\hbar$ where $E_0\sim 2 \, \rm{V/m}$ is the zero-point electric field in the cavity.   This yields a Jaynes-Cummings Hamiltonian of the joint system $H=H_e +H_r + \hbar g (a^{\dag} c+a c^{\dag})$, where $c$ is the motional quanta annihilation operator.  The electron motional states can be manipulated quickly due to the large coupling strength $g/2\pi=20\, \rm{MHz}$, a consequence of the large electron dipole moment $e a_x/\sqrt{2}  \sim 2 \times 10^3\, {\rm Debye}$, and without exciting transitions to higher lateral states due to the anharmonicity $\alpha/2\pi \approx -100 \, \rm{MHz}$.

In addition to the motional degree of freedom, the electron carries a spin degree of freedom.  The bare coupling of cavity photons to the spin is many orders of magnitude weaker than to the charge, but can be enhanced via controlled spin-motion coupling.  A different mechanism of enhancement for semiconducting double-dots was pointed out in~\cite{childress_mesoscopic_2004}. A spin-quantization axis is established using a magnetic field in the $\hat{x}$ direction (Fig.~\ref{fig:MotionalLevels}).  The Larmor frequency per unit field is approximately $\omega_{\rm L}/2\pi B = 2\mu_{\rm{B}}/h\approx2.89\, \rm{MHz}/\rm{G}$.  Niobium cavities have been demonstrated to maintain $Q>20,000$ in parallel fields of up to $2 \, \rm{kG}$, allowing Larmor frequencies of up to  $\omega_{\rm{L}}\sim 6 \, \rm{GHz}$.  Both the cavity and motion have $\hbar \omega \gg k_{\rm{B}} T$ so that they relax to the ground state.  

We propose to create a non-uniform $z$-field component with a gradient along the vibrational axis, $\partial_x B_z$, by passing a current through the center electrode (in the $y$-direction see Fig.~\ref{fig:MotionalLevels}). This leads to a new term in the Hamiltonian, $H_{\rm{s}}=-2\mu_{\rm{B}}s_z x \partial B_z/ \partial x$. The resulting spin-orbit interaction provides an enhanced cavity coupling $g_s$, mediated through the motional state, where
\begin{equation}
g_s=  \mu_{\rm B} a_x \frac{\partial B_z}{\partial x}  \frac{g \sqrt{2}}{\hbar \omega_x (1-\omega_L^2/\omega_x^2)}
\end{equation}
This allows manipulation and readout of individual spins, as well as the use of coupling techniques developed for superconducting qubits~\cite{dicarlo_demonstration_2009}.    Further, the coupling is proportional to the applied current, allowing the spin-cavity to be tuned in-situ on nanosecond timescales. For a 1 mA current 500 nm away a $\partial B_z/ \partial x \sim 8 \, \rm{mG}/\rm{nm}$ field gradient can be created.  If  $\omega_L \ll \omega_x$ these parameters give $g_s \sim 8 $~kHz whereas if $\omega_x - \omega_L \approx  30$~MHz then the coupling can be made large, $g_s \approx 0.5$~MHz.    

The current also creates a second-order variation in the $x$-component of $\vec{B}$, leading to a new term in the Hamiltonian, $H_{\rm sb} =-\mu_{\rm{B}}x^2\partial_x^2 B_x s_x$.  If the constant magnetic field is applied along the $y$-direction, this term will lead to sideband transitions simultaneously changing the orbital and spin states for drives at $\omega_{\pm}=\omega_{x}\pm\omega_{\rm{L}}$.    These transitions can be used to manipulate, cool, and detect the spin using its coupling to the lateral motion.  With such cooling it might allow one to use smaller spin frequencies.  

It is also important to consider decoherence of the motional and spin states.  The two major sources of noise are electrical fluctuations in the leads and excitations in the liquid helium.  Here we present a short summary of these decoherence mechanisms (also see Fig.~\ref{fig:Decoherence}).  A detailed explanation of these mechanisms is presented in the supplementary materials~\cite{eonhe_supplement}. 

A motionally excited electron can relax radiatively via spontaneous emission directly into free space, through the cavity, or the trap bias electrode.  The electron radiates very little into free space, both because it is small ($a_x \ll \lambda$), and because the microwave environment is carefully controlled.    In a perfectly symmetric trap, radiation through the bias leads would be suppressed by a parity-selection rule.  We conservatively assume that the electron is displaced from the trap center by $\sim a_x$, which gives a relaxation rate $\sim 1.6\times 10^3\, s^{-1}$.  Though this mechanism is not expected to be dominant, it could be easily reduced significantly by engineering the impedance of the trap bias lead.  In addition, slow fluctuations in the trap electrode voltage ($V_{\rm{e}}$) can deform the potential, changing the motional frequency and resulting in dephasing.  This can occur from drift in the voltage source, thermal Johnson voltage noise, or local ``1/f'' charge noise.  Drift slow compared with the experiment time is easily compensated.  The thermal noise at 50 mK is quite small with dephasing rate $<100$~Hz.  Any charge fluctuations in the bias leads should be screened by their large capacitance to ground but even conservatively assuming an anomolously small capacitance, we estimate a dephasing rate $8\times 10^{3}\, s^{-1}$, which would not be the dominant decoherence rate.  Noise from the cavity and ground plane electrodes should have less effect due to the symmetry of the potential.

\begin{figure}
\centering
\includegraphics{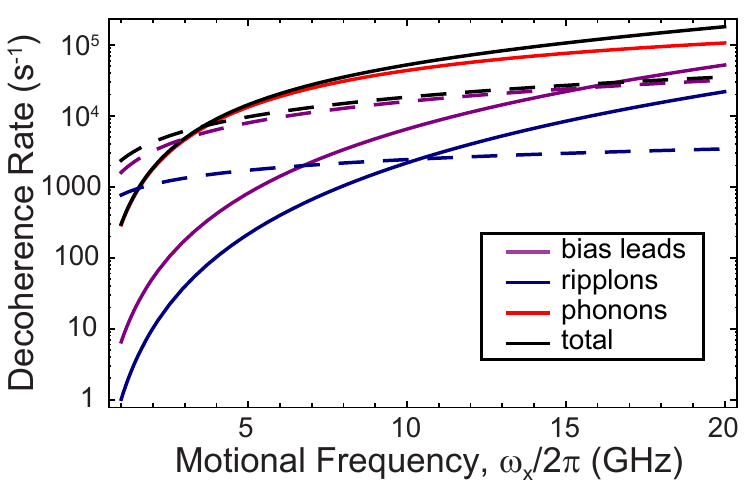}
\caption[Decoherence rates of electron motional state]{(color) Decoherence rates of motional states as a function of the trap frequency due to interactions with bias leads, ripplons, and phonons.  Rates are computed using parameters specified in the text at $T=50\,{\rm mk}$.  Solid lines are decoherence rates due to energy relaxation ($\Gamma_1/2$), while dashed lines are dephasing rates ($\Gamma_{\phi}$).  Single ripplon relaxation rate and phonon dephasing rates are $<1$ Hz.   Spin decoherence rates are discussed in the text. }\label{fig:Decoherence}
\end{figure}

In addition to decoherence through the electrodes, the electron can lose coherence to excitations in the helium. Two major types of excitations are relevant: capillary waves on the helium surface, known as ripplons, and phonons in the bulk. The electron is levitated above the surface at height $r_{\rm B}\sim 8$~nm, which greatly exceeds the height of the surface fluctuations, and therefore coupling to ripplons is small.  The characteristic electron speed $a_x\omega_x$ significantly exceeds the speed of sound $v_s$ in He and the characteristic group velocity of ripplons. As a result, the rate of direct emission is suppressed and decay into ripplons is dominated by second order processes in which two ripplons of nearly opposite momentum simultaneously interact with the electron.  The allowed phase volume is limited by the condition on the total ripplon momentum. Thus the corresponding decay rate is small, estimated to be $\lesssim 10^3$~s$^{-1}$ (see Fig.~\ref{fig:Decoherence}). 

The most important mechanism related to helium excitations is decay into phonons. The coupling to phonons is reminiscent of piezoelectric coupling in semiconductors. An electron creates an electric field that causes helium polarization, which in turn affects the electron. Phonons modulate the helium density and thus the polarization, which changes the electron energy. However, in contrast to semiconductors, where the typical piezoelectric constant is $e_{\rm pz}\sim 10^{14}$~$e$/cm$^2$~\cite{Seeger2004}, its analog in He is $\sim e(\epsilon -1)/4\pi r_{\rm B}^2 \sim 10^{10}~e/$cm$^2$. Therefore coupling to phonons is much weaker than in semiconductors. The corresponding decay rate is $\sim 3\times 10^4\, \rm{s}^{-1}$ (see Fig.~\ref{fig:Decoherence}). 

Besides decay, coupling to helium excitations leads to fluctuations of the electron frequency and ultimately to dephasing. The major contribution comes from two-ripplon processes, since ripplons are very soft excitations with comparatively large density of states at low energies, so that they are excited even for low temperatures. However, because of the weak coupling, the dephasing rate remains small, $\sim 2\times 10^3\, {\rm s}^{-1}$ for $T=50$~mK (see Fig.~\ref{fig:Decoherence}). It also decreases rapidly as the temperature is lowered. Another mechanism of dephasing are slow drifts in the helium film thickness, which change the trap frequency through its dependence on the height, $d$, of the electron.  Fortunately, the cavity forms a liquid He channel~\cite{glasson_observation_2001} in which the film height is stabilized by surface tension, rendering it much less susceptible to low frequency excitations.

The electron spin promises much longer coherence times, and when uncoupled to the charge the lifetime is expected to exceed seconds~\cite{lyon_spin-based_2006}.  When the spin is coupled to the motion, it will also inherit the orbital decoherence mechanisms with a matrix element  $\propto \mu_{\rm B}\partial_x B_z a_x /\hbar\omega_x$.  These mechanisms can be further diminished by turning off the gradient field or changing the spin-motion detuning, to reduce the coupling.  In addition to decoherence felt through the spin-orbit coupling, the electron spin can be dephased by fluctuating magnetic fields.  These can arise from Johnson current noise in the leads which would lead to dephasing rates less than $1\, s^{-1}$.  It is also possible that the spin will be affected by ``1/f'' flux noise~\cite{koch_flicker_1983}, often seen in SQUID experiments.  The trap involves no loops or Josephson junctions, so it is difficult to predict to the extent of flux noise in this geometry, however even a worst case estimate still yields a dephasing rate of only $200\,s^{-1}$~\cite{eonhe_supplement}.  

In summary we use circuit QED to propose solutions to many of the problems associated with electrons on helium, developing the ability to manipulate and detect both the electron's quantized motion and its spin.  Further, this architecture couples electrons on helium to each other and to other quantum systems via single microwave photons, creating a scalable architecture for quantum computing.  

The authors would like to acknowledge useful discussions with Forrest Bradbury.  This work was supported in part by the NSF grants DMR-053377, EMT/QIS 0829854 and CCF-0726490.  Institutional support was also provided by Yale University via the Quantum Information and Mesoscopic Physics Fellowship (DIS) and the Austrian Academy of Sciences through the DOC fellowship (AF).


%

\end{document}


\title{Supplementary material for ``Circuit Quantum Electrodynamics with Electrons on Helium''}
    \author{David Schuster}
    \affiliation{Department of Applied Physics and Physics, Yale
    University}
    \author{Andreas Fragner}
    \affiliation{Department of Applied Physics and Physics, Yale
    University}
    \author{Mark Dykman}
    \affiliation{Department of Physics and Astronomy, Michigan State University}
    \author{Steve Lyon}
    \affiliation{Department of Electrical Engineering, Princeton University}
    \author{Rob Schoelkopf}
    \affiliation{Department of Applied Physics and Physics, Yale
    University}
\date{\today}

\maketitle

\section{Introduction}

These supplementary materials estimate the rates of relaxation and dephasing of the quantized electron motion in a small electrostatic trap.  A variety of decoherence mechanisms are explored including, electric and magnetic noise from the electrodes as well as emission and scattering of ripplons and phonons.  We use DPS where we refer to Dykman et al, PRB {\bf 67}, 155402 (2003)~\cite{dykman_qubits_2003}. The notations below are from this paper. We refer to Eq.~(X) in this paper as DPS~(X).  A discussion of the relaxation and dephasing of the electron spin can be found elsewhere~\cite{lyon_spin-based_2006}.

\section{Definitions and useful Relations}

In addition to the constants defined in the main text and DPS there are some definitions and relations that will be useful throughout these materials.  The motional state of the electron is represented as $\ket{n_x,n_y}$ where $n_{\{x,y\}}$ is the quantum number in the $x,y$ direction. The corresponding transition frequencies are $\omega_{i}$ and the localization lengths are $a_i=(\hbar/m\omega_i)^{1/2}$, with $i=x,y$.  In this supplement the electron is assumed to remain in the ground state of motion normal to the helium surface.

With these notations, we have
\begin{eqnarray}
\label{eq:gl}
g_{l}(\textbf{q})&=&\left|\bra{1,0}e^{i \textbf{q r}}\ket{0,0}\right|^2 \nonumber \\
&=&\frac{1}{2}(q_x a_x)^2 e^{-\sum_j q_j^2 a_j^2/2},
\end{eqnarray}
\begin{eqnarray}\label{eq:gph}
g_{ph}(\textbf{q}) &=&\left|\bra{1,0} e^{i \textbf{q r}} \ket{1,0}-\bra{0,0} e^{i \textbf{q r}} \ket{0,0}\right|^2 \nonumber \\
&=&\frac{1}{4}(q_x a_x)^4 e^{-\sum_j q_j^2 a_j^2/2}.
\end{eqnarray}

Respectively,
\begin{eqnarray}
\frac{1}{(2 \pi)^2} \int d\textbf{q}g_{l}(\textbf{q}) &=& \frac{1}{4\pi a_x a_y} \nonumber\\
\frac{1}{(2 \pi)^2} \int d\textbf{q}g_{ph}(\textbf{q}) &=& \frac{3}{8\pi a_x a_y} \label{eq:gph2}
\end{eqnarray}

The kinetic energy normal to the helium surface for zero pressing field $\Ep$ is
\begin{eqnarray}
_z\bra{1}p_z^2/2m\ket{1}_z &=& R         \nonumber             \\
R&=&\frac{\hbar^2}{2 m r_{\rm{B}}^2} \approx h \times 158\, \rm{GHz} \nonumber                 \\
r_{\rm{B}}&=&\hbar^2/ \Lambda m \approx 7.64\, \rm{nm},         \nonumber             \\
\Lambda &=& \frac{(\epsilon_{\rm he}-1) e^2}{4 (\epsilon_{\rm he}+1) },
\end{eqnarray}
where $\ket{1}_z$ is the ground state wave function of motion normal to the helium surface.

It is also helpful to include several properties of helium.  The dielectric constant of helium is $\epsilon_{\rm he} \approx 1.057$.  The density of helium is $\rho = 0.145\, {\rm g/cm}^3$, with surface tension $\sigma = .378 \,{\rm dyne /cm}$, and a dispersion relation for capillary waves (ripplons) is $\omega_q = (\sigma q ^3/\rho)^{1/2}$.  For phonons the dispersion relation is $\omega_Q = Q \nu_s$, with $\nu_{\rm s} \approx 2.4\times 10^4 {\rm cm/s}$ the speed of sound in superfluid helium.

For obtaining numerical results the following discussion will assume that the electron vibrational frequency is $\omega_x \approx 5\,\rm{GHz}$, the zero-point motion of the electron is $a_x \approx 4.8\, \rm{nm}$, and $g\approx 25\, \rm{MHz}$.

\section{Lifetime}

The general prescription of this section will be to write down Fermi's golden rule with the appropriate matrix element for the specific process.  The matrix element will then be evaluated and integrated over the appropriate density of states.

\subsection{Photon emission}

A vibrating electron can emit a photon via electric-dipole radiation into the vacuum.  If the electron were in vacuum it would radiate at a rate
\begin{equation}
\Gamma_1^{\rm (v)}= \frac{2}{3} \frac{e^2}{\hbar c} \left(\frac{2\pi a_x}{\lambda}\right)^2 \omega_x
\end{equation}
where $\lambda=2\pi c/\omega_x$ is the wavelength of the vibrational frequency.  If this were the only relaxation process the excited state would last for longer than 100 s.

Of course, the electron is embedded in the electromagnetic environment created by the microwave resonator and bias wiring.  For a simple two-level dipole coupled to a single mode cavity with frequency $\omega_c$ for nonresonant coupling, $g \ll |\omega_x-\omega_c|$, the spontaneous emission rate is
\begin{equation}
\Gamma_1^{\rm (v)} = \frac{g^2\kappa}{(\omega_x-\omega_c)^2}
\end{equation}
where $\kappa$ is the mode decay rate. In the resonant case and in the bad cavity limit, where the cavity decay is faster than the coupling and the detuning, $\kappa \gg g, |\omega_x-\omega_c|$, the decay rate is $\Gamma_{v}=g^2/\kappa$.  In the strong coupling limit, the rate of emission is $\Gamma_1^{\rm (v)}=\kappa/2$ on resonance.  It is also possible to model the case of a dipole in multi-mode cavity circuit~\cite{houck_controllingspontaneous_2008}.

In addition to the emission through the cavity, the electron can also decay through the trap bias lead.  This decay rate can be found by considering the effect of the Nyquist noise, $S_{V_e} (\omega_x) \approx 2 \hbar \omega_x {\rm Re}[Z(\omega_x)]$ of the bias electrode on the electron.   If there are no other fields, the bias electrode only couples to $\hat{x}^2$, and would not cause relaxation.  However any displacement $\Delta x$ from the bias electrode null, due to stray or intential DC fields from other electrodes will add a $\hat{x}$ interaction and will open a channel of relaxation.  In this case the coupling to the electrode potential fluctuations $\delta V_e$ is given by $\hat{h} \delta V_e$, with $\hat{h}=e \hat{x} \partial E_x/\partial V_e$, where $E_x$ is the electric field on the electron in the $x$ direction due to the bias electrode.  The decay rate is then
\begin{equation}\label{eq:GammaBias1}
\Gamma_1^{\rm (el)} = \frac{1}{\hbar^2} \left|\bra{0,0}\hat{h}\ket{1,0}\right|^2 S_{V_e}(\omega_x).
\end{equation}
For small $\Delta x$ compared to the inter-electrode distance, we have $e E_x = m_e \omega_x^2 \Delta x = \hbar \omega_x \Delta x/ a_x^2$.   Substituting these results into Eq.~\ref{eq:GammaBias1} and allowing for $\partial E_x/\partial V_e = E_x/V_e$, one obtains
\begin{equation}
\Gamma_1^{\rm (el)} = \frac{{\rm Re}[Z(\omega_x)]}{\hbar/e^2} \left(\frac{\hbar \omega_x}{e V_e}\right)^2 \omega_x
\end{equation}
Even assuming that the electron is displaced significantly, $\Delta x \approx a_x$ and ${\rm Re}[Z(\omega_x)]\approx 50\, \Omega$, this relaxation is quite small $\Gamma_1^{\rm (v)} \approx 1.6 \times 10^{3} s^{-1}$ and should not be a limiting factor.  If this becomes a hindrance in the future it can be reduced by several orders of magnitude by changing ${\rm Re}[Z(\omega_x)]$ using a resonant structure.

\subsection{Two-ripplon scattering}

Single ripplon scattering is exponentially suppressed by the mismatch of the size of the electron wave function to the ripplon wavelength at the same energy.  The dominant decay process is one in which two ripplons are emitted in nearly opposite directions, each with approximately half of the electron energy $\hbar\omega_x$.

The decay rate due to such a process can be estimated by applying Fermi's Golden rule to DPS(11)
\begin{eqnarray}\label{eq:FGH2Ripplon}
\Gamma_1^{(2\rm r)} = &&\frac{2 \pi}{\hbar} \sum_{\qb_1,\qb_2} \left|\bra{0,0}\xi_{\qb_1} \xi_{\qb_2} e^{i (\qb_1+\qb_2)\cdot \rb} V_{\qb_1,\qb_2}\ket{1,0}\right|^2  \nonumber \\
&& \times \delta (\hbar \omega - \hbar \omega_{q_1}-\hbar \omega_{q_1}) (\bar n_{q_1}+\bar n_{q_2}+1),
\end{eqnarray}
where $\xi_{\qb}$ is the helium surface displacement due to a ripplon with wave vector $\qb$ and $\bar n_q\equiv \bar n(\omega_q)$, with $\bar n (\omega)=\left[\exp(\hbar\omega/k_{\rm B}T)-1\right]^{-1}$.
This calculation can be simplified significantly by noticing that, because of the Gaussian form of the matrix element Eq.~(\ref{eq:gl}), $|\qb_1+\qb_2| \leq 1/a_x$.  Since $q_{\rm{res}}$ found from condition $2\omega_{q_{\rm res}}=\omega_x$ is $\approx 4.6 \times 10^6 \, \rm{cm}$ and $q_{\rm{res}} a_x \approx 30$, we approximate $\left|\qb_1\right|\approx\left|\qb_2\right|=q_{\rm res}$.

The strength of the direct two-ripplon coupling $V_{\qb_1\qb_2}$ is a sum of contributions from the kinetic, or inertial term (from electron accommodating to the curvature of the barrier on the helium surface), and a polarization term, which comes from the change of the image potential due to the curvature of the surface.  In Eq.~\ref{eq:FGH2Ripplon} and below it should be assumed unless otherwise stated that $V_{\qb_1\qb_2}$ is the matrix element of the interaction with ripplons or phonons on out-of plane wave functions $\ket{1}_z$. A simple calculation shows that the main contribution to scattering comes from the kinetic term, see DPS~(11)-(13),
\begin{eqnarray}
\label{eq:two_rippl}
V_{\qb_1\qb_2} \approx -R\qb_1\qb_2.
\end{eqnarray}
A correction from the polarization term, DPS~(13), is comparatively small. The corresponding matrix element of the interaction is
\begin{equation}
\Lambda z^{-3}[2-x^2K_2(x)] \approx 4.6 R/r_B^2
\end{equation}
with $x=q_{\rm res}z$.  The ratio of this term to the kinetic-energy term is 40\%. The remaining term in the polarization coupling, DPS~(13), is even smaller and partly compensates the above term. The contribution of the polarization coupling decreases, relatively, when we apply a pressing field.

Using this approximation to evaluate the matrix element and converting the sum to an integral yields
\begin{eqnarray}
&&\Gamma_1^{(2\rm r)} = \frac{\hbar}{16 \pi^2 \rho^2} \left(\int d\textbf{q}g_{l}(\textbf{q})\right) \nonumber\\
&&\times\int dq \frac{q^3}{\omega_q^2} \left(q^2 R\right)^2 \delta (\hbar \omega-2 \hbar \omega_{q}) (2 \bar n_{q}+1) \nonumber\\
&&\qquad \approx \frac{R^2}{192\pi a_xa_y}\frac{\rho^{2/3}\omega_x^{7/3}}{2^{1/3}\sigma^{8/3}}
\end{eqnarray}
This rate is small, $\sim 450$~s$^{-1}$.

\subsubsection{Renormalization due to single-ripplon coupling}

The single-ripplon coupling leads to renormalization of the two-ripplon coupling. The renormalization leads to the replacement of the matrix element of two-ripplon coupling
\begin{eqnarray}
\label{eq:renormalization}
&&\bra{\mu} e^{i(\qb_1+\qb_2)\rb}\hat V_{\qb_1\qb_2}\ket{\nu}
 \to \bra{\mu}e^{i(\qb_1+\qb_2)\rb}\hat V_{\qb_1\qb_2}\ket{\nu} \nonumber\\
&&- \frac{1}{2}\sum_{\kappa}V_{\mu\kappa}(\qb_1)V_{\kappa\nu}(\qb_2)
\left[(\ep_{\kappa}-\ep_{\nu})^{-1} + (\ep_{\kappa}-\ep_{\mu})^{-1}\right], \nonumber \\
&&V_{\mu\nu}(\qb)=\langle\mu|e^{i\qb\rb}\hat V_{\qb}|\nu\rangle.
\end{eqnarray}
Here, $|\mu\rangle$ are electron states ($\mu$ enumerates both in-plane and out-of-plane states), $\ep_{\mu}$  are the state energies. We have disregarded the dynamics of ripplons, i.e., we assume that the change of the electron energy in a virtual transition is much larger than the ripplon energy change. The explicit form of the coupling matrix element (an operator with respect to motion normal to the surface) $\hat V_{\qb}$ is given by DPS~(10). We need virtual transitions into such states $\mu$ that the matrix elements $\langle \mu_{\parallel}|\exp(i{\bf q r})|0,0\rangle$ is of order 1 for $q\sim q_{\rm res}$, where $|\mu_{\parallel}\rangle$ is the in-plane component of the wave function $|\mu\rangle$. This condition leads to typical $\ep_{\mu}\approx \hbar^2q_{\rm res}^2/2m$. Such energy is very large, $\sim 12 R$. The terms in $\hat V_{\qb}$ that are $\propto q^2$ are also large for such $q$. They lead to renormalization~\cite{konstant_in_preparation} of the matrix elements of direct two-ripplon coupling, reducing them by a factor $\propto (qr_B)^{-1}$ which is $\sim 0.3$ for $q=q_{\rm res}$. The polarization coupling is also reduced. Therefore, the rate of two-ripplon decay is reduced down to $< 10^3$~s$^{-1}$.

\subsection{Phonon scattering}

Another way that the electron motion can relax is through phonon emission.  The electron can launch a phonon by two mechanisms.  In the first the coupling is mediated by modulation of the dielectric constant along with the density wave.   In the second, the electron can couple to the surface displacement in much the same way as in the ripplon case but now launching most of the energy normal to the surface.

\subsubsection{Phonon scattering: dielectric constant modulation}

For frequency $\omega_x=5$~GHz and phonon sound velocity $\nu_s=2.4\times10^4$~cm/s, we have the normal to the surface component of the wave vector of the resonant phonon $(Q_z)_r=\omega_x/\nu_s\approx 1.3\times10^6$~cm$^{-1}$, whereas the typical in-plane wave number $q_x=(m\omega_x/\hbar)^{1/2}\approx 1.6\times 10^5$~cm$^{-1}$. This shows that the phonons involved in inelastic scattering propagate almost normal to the surface.  We will modify DPS~(25) - (27) to write the interaction in terms of the velocity potential
\begin{eqnarray}
\label{eq:velocity_potential}
&&\phi(\rb,z)=\sum_{\Qb} \phi_{\Qb}\left(\hat c_{\qb \,Q_z} + \hat c^{\dagger}_{-\qb\,Q_z}\right)e^{i\qb\rb}\sin(Q_zz), \nonumber\\
&&\phi_{\Qb}= (\hbar \nu_s/\rho V Q)^{1/2}, \qquad \Qb=(\qb,Q_z),
\end{eqnarray}
where $\hat c_{\qb\,Q_z}$ is the annihilation operator of a phonon with the wave number $(\qb, Q_z$); $\rb,\qb$ are two-dimensional vectors and $Q_z>0$. Using the Fermi golden rule, for $\bar n(\omega_x)\ll 1$ the decay rate can be written as
\begin{eqnarray}
\label{eq:pol_phonons_general}
\Gamma^{\rm (d)}_{1} &=& \frac{2 \pi}{\hbar} \sum_{\Qb} \left|\bra{0,0} \phi_{\Qb} e^{i \textbf{q}\cdot\textbf{r}} V_{\Qb}^{\rm (d)}\ket{1,0}\right|^2 \nonumber \\
&& \times\delta (\hbar \omega_x - \hbar \omega_Q^{\rm (ph)}), \\
V_{\Qb}^{\rm (d)} &=&\frac{ i \Lambda q Q}{\nu_s}\, _z\bra{1}v^{\rm (d)}\ket{1}_z, \nonumber \\
v^{\rm (d)} &=& \int^{\infty}_0 dz' (z+z')^{-1} \sin( Q_z z') K_1 [q (z+z')].\nonumber
\end{eqnarray}
Here, $\omega_Q^{\rm (ph)}$ is the phonon frequency.

Using the explicit form of the ground-state wave function and taking into consideration that the typical in-plane wave numbers are small, $q\ll 1/r_B$, so that in Eq.~(\ref{eq:pol_phonons_general}) $K_1[q(z+z')]\approx 1/q(z+z')$ (this is an overestimate), one can write the expression for the decay rate due to modulation of the dielectric constant $\ep$ as
\begin{eqnarray}
\label{eq:decay_ph_pol}
&&\Gamma^{\rm (d)}_{1}=32R^2\omega_x/\left(\pi \rho \nu_s^3\hbar a_xa_y\right)\nonumber\\
&&\times \left|\int\nolimits_0^{\infty}dz_1\,dz_2\left(\frac{z_1}{z_1+z_2}\right)^2
e^{-2z_1}\sin(Q_z r_B z_2)\right|^2
\end{eqnarray}

The dimensional factor in front of the integral in Eq.~(\ref{eq:decay_ph_pol}) is  $ 4.3\times10^6$~s$^{-1}$. With $(Q_z)_rr_B^{(0)}=1.01$ this gives the overall rate of $\sim 2.7\times10^4$~s$^{-1}$.

\subsubsection{Phonon scattering: surface displacement}

The contribution to the decay rate comes also from the displacement of the helium surface, which can be considered a free surface for high-energy phonons propagating almost normal to it. This is much like the ripplon case, but here a single-phonon decay is allowed because the 2D momentum is conserved while the excess energy is dumped into the normal component.  The decay rate can be expressed as
\begin{eqnarray}
\Gamma^{\rm (s)}_{1}=&&\frac{2 \pi}{\hbar \nu_s^2} \sum_{\Qb} \left|\bra{0,0}\phi_{\Qb} e^{i \textbf{q} \cdot \textbf{r}} \overline{V_{\qb}}\ket{1,0}\right|^2 \nonumber\\
&&\times\delta (\hbar \omega - \hbar \omega_Q^{\rm (ph)}).
\end{eqnarray}
Here, $\overline{V_{\qb}}$ is the diagonal matrix element of the coupling operator $\hat V_{\qb}$, which is detailed in DPS~(9) - (10), calculated on the wave functions of the ground state of motion normal to the helium surface,
\begin{equation}
\label{eq:one_vibration_coupling}
\overline{V_{\qb}}={_z\bra{1}}\hat V_{\qb}\ket{1}_z=e\Ep+ \Lambda q^2\langle v_{\rm pol}\rangle,
\end{equation}
where $\Ep$ is the electric field that presses the electron against the surface and $\langle v_{\rm pol}\rangle = {_z\bra{1}}(qz)^{-2}[1-qzK_1(qz)\ket{1}_z$. Then
\begin{eqnarray}
\Gamma_{1}^{\rm(s)}=&&\frac{2}{\hbar \rho \nu_s \omega_x} \frac{1}{(2\pi)^2}\int d\textbf{q} g_{l}(\textbf{q})\left|\overline{V_{\qb}}\right|^2
\end{eqnarray}

The contribution from the pressing field is
\begin{eqnarray}
\label{eq:decay_ph_displac_eperp}
\Gamma_{1}^{(\Ep)}=\frac{e^2\Ep^2}{2\pi\hbar\rho \nu_s\omega_xa_xa_y}
\end{eqnarray}
For a comparatively strong field $\Ep=300$~V/cm this rate is $8.6\times10^3$~s$^{-1}$.

We now consider the contribution to $\Gamma_{1}^{\rm(s)}$ of the term $\langle v_{\rm pol}\rangle$. For typical $q\lesssim 1/a_x$ we have  $\langle v_{\rm pol}\rangle \sim -\ln(qr_B)/2$. Using this approximation, we obtain for the corresponding contribution
\begin{eqnarray}
\label{eq:polar_sound}
\Gamma_{1}^{(\rm pol)} &=& \frac{\Lambda^2}{2\pi^2\hbar\rho \nu_s\omega_x}\int d\qb \,q^4g_l(\qb)|\langle\nu_{\rm pol}\rangle|^2\nonumber\\
&&\sim \frac{48 R^2r_B^2}{\pi\hbar\rho \nu_s\omega_x a_x^6}\left[\ln(r_B/a_x)/2\right]^2.
\end{eqnarray}
Here, we assumed that $a_x\sim a_y$. The last factor is somewhat smaller than the numerical result for actual $r_B/a_x$. For a numerical estimate we set it equal to one. This gives $1.8\times10^3$~s$^{-1}$. There is an interference term of the two last expressions, but it is smaller than their sum.

Therefore, polarization phonon scattering is the dominating mechanism of inelastic scattering due to excitations in liquid helium.

\subsection{Lifetime summary}

There are several contributions to the lifetime of the excited vibrational state of the electron.  There are effects which could modify these results (though the estimates here are conservative), but in the current analysis the dominant mechanism appears to be phonon emission, coupled mostly via modulation of the dielectric constant. It gives the lifetime of the vibrational state $T_1 \approx 35 \rm{\mu s}$.

\section{Dephasing}

In addition to loss of coherence of the electron due to decay processes, excitations in the environment at low frequencies can cause variations in the transition frequency, leading to dephasing of superposition states.  In this section we treat dephasing from voltage noise, variations in helium level, and state dependent quasi-elastic scattering of ripplons off the electron.

\subsection{Voltage noise}

The depth of the potential and thus the electron transition frequency depends on the trap bias voltage.  Any low frequency noise on the bias electrode will result in electron phase noise. Noise on the bias lead can arise from slow drift of the voltage source, thermal fluctuations (Johnson noise), or anomalous local sources ($1/f$ noise).  For typical electrode geometries the transition frequency scales like $\omega_x \propto V_e^{1/2}$.  Small changes in $V_e$ cause a frequency shift $\delta \omega = -\omega_x \delta V_e /2 V_e$.  Typical long term drifts for precision voltage sources are $\delta V_e/V_e\approx 10^{-6}$, or a drift of about $\delta \omega_x\approx 8\times 10^3 s^{-1}$.  Because this drift is typically on hour timescales it should be easily compensated by measuring the transition frequency and readjusting the voltage. For thermal noise~\cite{martinis_decoherence_2003}
\begin{eqnarray}\label{eq:GammaPhiVeWhite}
\Gamma_{\phi}^{(v)} &=& \left(\frac{\partial \omega_x}{\partial V_e}\right)^2 S_{V_e}/2  \nonumber \\
&=& \frac{\omega^2 k_{\rm{B}} T Re\left[Z(\omega)\right]}{4 V_e^2}
\end{eqnarray}
If superconducting leads are used then there is no local Johnson noise but noise can still be coupled in from dissipative sources off chip.  For the simulated bias voltage $V_e \approx 10\, \rm{mV}$ and conservatively assuming an environmental impedance of $Z=50\, \Omega$, Eq.~\ref{eq:GammaPhiVeWhite} gives $\Gamma_{\phi}^{(v)} \approx 90 s^{-1}$.

Usually the dominant source of electrical dephasing is $1/f$ noise which is thought to arise from mobile charges in the substrate.  This type of noise spectrum does not lead to a simple dephasing rate. The diffusion of phase can be estimated as~\cite{martinis_decoherence_2003}
\begin{eqnarray}
\langle[\phi(t)-\phi(0)]^2\rangle &\sim & \left(\frac{\partial \omega_x}{\partial V_e}\right)^2 S_{V_e} (1\,\rm{Hz}) \ln (0.4/f_m t) t^2 \nn \\
&=& \frac{(\omega_x t)^2}{4 V_e^2} S_{V_e} (1\,\rm{Hz}) \ln (0.4/f_m t)
\end{eqnarray}
where $f_m^{-1}$ is the total averaging time, and $t$ is the precession time of single measurement.  The dephasing time for uncompensated $1/f$ noise can be estimated as the time for $\langle[\phi(t_\pi)-\phi(0)]^2\rangle = \pi$.  Local charges should be screened by the bias electrode.  To get an estimate of their effect we assume that $S_{V_e} = S_q / C_{\rm eff}$ and that the effective capacitance is only $C_{\rm eff} \approx 1\, \rm{fF}$ (though it should be much more than that) and the ``typical'' charge noise of $S_q(1\, \rm{Hz}) \approx 10^{-4}\, \rm{e}/Hz^{1/2}$.  With these assumptions $t_{\pi}^{-1} = 8\times 10^3\, \rm{s}^{-1}$.  Even these conservative assumptions lead to rates smaller than the relaxation rates, which can be seen as a result of the electron being physically separated from the electrodes where the charges may fluctuate.

\subsection{Fluctuations in helium level}

The potential energy landscape is created by electrostatic gates beneath the helium surface.  Any fluctuations in the thickness of the helium above the electrodes will change the effective voltage seen by the electron resulting in changes to the transition frequency.  While quantized excitations in the helium level will be treated in the next section, here we explore the effect and susceptibility to slow changes in the level of liquid helium in the reservoir due to fluctuations in temperature, or external vibrations.  This type of fluctuation is unlikely to cause dephasing during the lifetime of the qubit and should be susceptible to spin-echo techniques, but if unchecked can lead to slow drifts in the frequency, making it difficult to bias the electron reliably.

For the geometry simulated in Fig. 2 the transition frequency has a sensitivity to small fluctuations in the mean helium film thickness of $\partial \omega_x/\partial d \approx 10\, {\rm MHz}/{\rm nm}$.  Fortunately superfluid has two properties, the formation of a Van der Waals film, and ideal capillary action that help stabilize the film thickness.  Exploiting these properties, the electron can be trapped above a channel~\cite{marty_stability_1986} in which the thickness is determined by the geometry and surface tension.  The trap is supplied by the Van der Waals creep film from a reservoir located well below the trap~\cite{crum_superfluid_1974}, significantly reducing its sensitivity to fluctuations in the reservoir height.  To accomplish this we design the guard ring of the trap to be much thicker than the bias electrode, forming a channel of the desired height.    The superfluid obeys Jurin's law, $R_{\rm c}=2\sigma/ \rho g H$ causing it to fill the trap if it is smaller than $R_{\rm c}$, the capillary radius of curvature~\cite{marty_stability_1986}.  For $H=5\,\rm{mm}$, the capillary length is $R_{\rm c}\approx 53 \, \rm{\mu m}$ and the trap is well filled, with sensitivity reduced by $\partial_H R_{\rm c} \partial_{R_{\rm c}} d = w/H  \approx 10^{-4}$.

\subsection{Direct two-ripplon coupling}\label{sec:rippDephasing}

An important contribution to dephasing is due to thermally excited ripplons. For T= 100 mK the typical wave number of a thermal ripplon ($\hbar\omega_{q_T}=k_{\rm B} T$) is $q_T\approx 4.1\times 10^6$ cm$^{-1}$. Therefore $q_Tr_B \approx 3.1$. For T=50 mK we have $q_T\approx 2.5\times 10^8$~cm$^{-1}$ and  $q_T r_B\approx 1.96$). The dephasing rate is given by the difference of the diagonal interaction matrix elements on the wave functions of the excited and ground vibrational states. The overall interaction matrix elements need to be projected on the ground state of out-of-plane motion, that is the relevant interaction can be written as [cf. DPS~(32)]
\begin{eqnarray}
\label{eq:dephase_inter}
H_i^{\phi}=\sum_{j=1,2}\sum_{\qb_1,\qb_2} v_j(\qb_1,\qb_2)b_{\qb_1}^{\dagger}b_{\qb_2}|j,0\rangle\langle j,0|
\end{eqnarray}
with
\begin{eqnarray*}
v_j(\qb_1,\qb_2)=&&(\hbar/\rho)(q_1q_2/\omega_{q_1}\omega_{q_2})^{1/2}
 V_{\qb_1\,-\qb_2}\\
&&\times\bra{j,0}\exp[i(\qb_1-\qb_2)\rb]\ket{j,0}.
\end{eqnarray*}

A useful relation for the following calculation is Eq.~(\ref{eq:gph}).
For $q_Tr_B\gg 1$ the major direct two-ripplon coupling is the kinematic coupling, with the projection given by Eq.~(\ref{eq:two_rippl}). With account taken of DPS~(32)-(33), the dephasing rate given by this interaction has the form
\begin{eqnarray}
\label{eq:two_rippl_dephase_general_kinetic}
&&\Gamma_{\phi}^{\rm (2r)} = \nonumber \\
&&\frac{\pi R^2}{S^2}\sum\nolimits_{\qb_1,\qb_2} \left(\frac{q_1}{\rho\omega(q_1)}\right)^2(\qb_1\qb_2)^2g_{ph}(\qb_1+\qb_2) \nonumber\\
&&\times\bar n_{q_1}\left(\bar n_{q_2}+1\right)\delta\left(\omega_{q_1}-\omega_{q_2}\right)
\end{eqnarray}
($S$ is the area). For $q_T\gg 1/a_{x,y}$, we can calculate this expression using that $|\qb_1-\qb_2|\ll q_{1,2}\sim q_T$. The calculation is further simplified if we set $a_x=a_y$, in which case we can average over the directions of $\qb_{1,2}$, which leads to the replacement $g_{ph}(\qb) \to (3/32)(qa_x)^4\exp(-q^2a_x^2/2)$. We can then integrate first over $\qb_2$ for given $\qb_1$ by writing $\qb_2= q'\hat \qb_1+ q''(\hat\qb_1\times\hat {\bf z})$ (hat indicates a unit vector). The inequality $|\qb_1-\qb_2|\ll q_{1,2}$ shows that $|q'|\gg |q''|$. To leading order in $(a_x q_T)^{-1}$ we have $\omega_{q_2}\approx \omega_{q'}$ and $\qb_1-\qb_2 \approx q''(\hat\qb_1\times\hat {\bf z})$. Integration over $q', q''$ then becomes simple, and the remaining integration over $\qb_1$ is then reduced to integration of the weighted temperature-dependent factor $\bar n_{q_1}(\bar n_{q_1}+1)$. The overall result is
\begin{eqnarray}
\label{eq:two_rippl_dephase_direct_kinetic}
\Gamma_{\phi}^{\rm (2r)}\approx \frac{(2\pi)^{1/2}}{192}\frac{\rho R^2}{\sigma^3a_{x}}(k_{\rm B}T/\hbar)^{3}.
\end{eqnarray}
For 100~mK this gives $\sim 1.4\times 10^4$~s$^{-1}$.

For T$\sim 100$~mK the polarization part of the direct two-ripplon coupling is small compared to the kinematic part. However, its role increases with decreasing temperature, since the characteristic ripplon wave number $q_T$ decreases. For $\qb_1\approx \qb_2=\qb$, the matrix element of the polarization coupling $-{_z\bra{1}}\Lambda z^{-3}[2-(qz)^2K_2(qz)]\ket{1}_z$ changes from $\approx  -0.9Rq^2$ to $\approx -0.6 Rq^2$ for $q$ increasing from $0.3/r_B$ to $1.2/r_B$. It is thus comparable with the kinematic coupling $Rq^2$ and has opposite sign. Therefore the overall contribution to the dephasing rate from direct two-ripplon coupling is significantly less than from the kinematic coupling taken alone, which from Eq.~(\ref{eq:two_rippl_dephase_direct_kinetic}) is $\sim 1.7\times 10^{-3}~{\rm s}^{-1}$ for T=50~mK.

\subsubsection{Renormalization due to single-ripplon coupling}

The kinematic and polarization single-ripplon coupling taken to the second order of the perturbation theory renormalizes the matrix elements of the direct two-ripplon coupling. For $q_{1,2}r_B\gg 1$ it reduces these matrix elements by a factor $\sim1/q_{1,2}r_B$~\cite{konstant_in_preparation}. This leads to a significant reduction of the dephasing rate for T=100~mK, since $q_Tr_B\approx 3$ for such temperature. On the other hand, for lower temperatures down to T=50~mK, where $q_Tr_B\approx 1.96$, the compensation is not strong, but the dephasing rate is already small here. The effect of the pressing field $\Ep$ remains smaller than from the polarization single-ripplon coupling for $\Ep\ll 10^3$~V/cm.

\subsection{Dephasing summary}

Ideally pure dephasing would be limited by the interaction of the electron motion with thermal ripplons.  This would set a dephasing time of $T_\phi \approx 0.5 \, \rm{ ms}$ even for T=100~mK.  This value would be very sensitive to temperature and could be reduced significantly at temperatures lower than 50 mK, giving it a distinctive signature.  It is possible, perhaps likely, that initial experiments will be dominated by anomalously high contributions from outside sources such as helium level fluctuations due to vibrations or spurious noise on the bias line, but given the analysis presented here it appears that it will be possible to eventually reduce these to acceptable levels.

\section {Spin Dynamics}
\subsection{Coupling to electromagnetic field}

The out-of-plane magnetic field $B_z$ that depends on the coordinates of in-plane motion (see Fig.~2) gives rise to spin-orbit coupling with Hamiltonian $H_{\rm{s}}=-2\mu_{\rm{B}}s_z x \partial B_z/ \partial x$. Since the orbital motion is coupled to the cavity mode, with Hamiltonian $H_{\rm g}=-ex\hat E$ ($\hat E$ is the operator of the electric field of the mode), the interaction $H_{\rm sb}$ leads to a spin-mode coupling mediated by lateral electron vibrations. Formally, for weak coupling, this interaction is described by the spin-mode vertex calculated in one electron-vibration line approximation. It can be easily found also without diagrams.

We will assume that the spin-orbit coupling is nonresonant, so that $|\omega_x-\omega_L|$ largely exceeds all relaxation rates and the detuning of the cavity mode from the Larmor frequency. A simple way to describe the spin-mode coupling is based on the equation of motion for operator $x$ in the Heisenberg representation. If we disregard the spin-orbit coupling, this equation reads
\begin{equation}
\label{eq:forced_vibrations}
m\ddot x +m\omega_x^2x=e\hat E.
\end{equation}
We seek the solution that describes forced vibrations at the bare Larmor frequency $\omega_L$, which is very close to the cavity mode frequency. Then we obtain $x\approx -e\hat E/[m(\omega_L^2 -\omega_x^2)]$. This solution is substituted into $H_{\rm{s}}$ instead of $x$, which leads to the spin-mode coupling $\propto s_z\hat E$. The coupling constant $g_s$ is given by Eq.~(2) of the main text. We note that the spin-orbit coupling causes also renormalization of the Larmor frequency $\propto (\mu_{\rm{B}}a_x\partial B_z/ \partial x)^2\hbar^{-2}\omega_L/(\omega_L^2-\omega_0^2)$.

\subsection{Spin decoherence}

In the absence of the enhanced spin-orbit interaction the lifetime of electron spin states is expected to exceed seconds~\cite{lyon_spin-based_2006}.  When the spin is coupled to the motion, it will also inherit the orbital decoherence mechanisms studied above with a matrix element reduced by $\propto \mu_{\rm B}\partial_x B_z a_x /\hbar\omega_x$.  These mechanisms can be reduced by turning off the gradient field or changing the spin-motion detuning, to reduce the coupling.  In addition to decoherence felt through the spin-orbit coupling, the electron spin can be dephased by fluctuating magnetic fields.  One source of magnetic field noise is Johnson (current) noise in the trap electrodes, a white noise given by $S_{\rm{I}}=4 k T  \Delta\omega / \rm{Re}~Z$, where $\Delta\omega$ is the bandwidth of interest, and the real part of the impedance is typically $\rm{Re}~Z\sim50 \, \Omega$.  The field created at the electron by a current in the wire is $B=4\times10^3\, \rm{mG}/\rm{mA}$, corresponding to a frequency shift of $\delta \omega_{\rm{L}}=3.50 \times 10^7\rm{s^{-1}}/\rm{mA}$.  The Johnson noise is extremely small $100\, \rm{pA}/\rm{Hz}^{1/2}$, giving a coherence time $T_{\phi,\rm{I}}\sim 20\, \rm{s}$.  Another possible source of magnetic field noise is $1/f$ flux noise~\cite{koch_flicker_1983}.   often seen in SQUID experiments with magnitudes of order $S_{\phi,\Phi}=10^{-6}\, \Phi_0 /\rm{Hz}^{1/2}$.   The trap geometry contains no superconducting loops or Josephson junctions so this mechanism may not cause decoherence for an electron on helium.  If present it is reasonable to assume the flux would be distributed evenly over the trap area ($\sim500\times500\, \rm{nm}^2$).  Even in this conservative scenario the dephasing rate would be $\Gamma_{\phi,\Phi} \sim 200 \, \rm{s}^{-1}$, allowing many coherent operations.
